\documentclass {llncs}
\usepackage{amssymb}
\usepackage{epsfig}
 \usepackage{url}
\usepackage{epsfig}
\usepackage{amssymb}
\usepackage{acronym}
 \usepackage{url}
 \usepackage{multirow}
 \usepackage[all]{xy}
   \usepackage{hyperref}
 \usepackage{amsmath}
 \usepackage{float}
  \usepackage{url}
  \usepackage{amssymb,amsmath,xspace,graphicx,multirow}
\restylefloat{table}
 
\usepackage{multirow}
\usepackage[all]{xy}
\usepackage{amsmath}
\begin{document}

\pagestyle{empty}

\mainmatter

\title{Towards the Formalization of Fractional Calculus in Higher-Order Logic\thanks{This is authors' version of CICM-2015 paper. The final publication is available at \url{http://link.springer.com}}}
\author{Umair Siddique* \and Osman Hasan** \and Sofi\`{e}ne Tahar*}
\institute{*Department of Electrical and Computer Engineering,\\ Concordia University, Montreal, Quebec, Canada \\
**School of Electrical Engineering and Computer Science, \\
National University of Sciences and Technology, Islamabad, Pakistan\\
\email{\{muh\_sidd,tahar\}@ece.concordia.ca}\\
\email{osman.hasan@seecs.nust.edu.pk}
\url{http://save.seecs.nust.edu.pk/projects/fc.html}} \maketitle

\begin{abstract}
Fractional calculus is a generalization of classical theories of integration
and differentiation to arbitrary order (i.e., real or complex numbers).
In the last two decades, this new mathematical modeling approach has been
widely used to
analyze a wide class of physical systems in various fields of science and
engineering. In this paper, we describe an ongoing project which aims at
formalizing the basic theories of fractional calculus in the HOL Light theorem
prover. Mainly, we present the motivation and application of such formalization
efforts, a roadmap to achieve our goals, current status of the project and future
milestones.
\end{abstract}

\section{Motivation and Background}
Physical and engineering systems are classified as continuous, discrete or hybrid depending upon the
nature of underlying system parameters. The rich theories of mathematics provide the necessary tools to study the
behaviour of such systems ranging from very small biological organisms to the modern Quantum mechanical phenomenons.
Generally, differential equations \cite{math_modeling} and difference
equations \cite{diff_eq_book} are used to characterize the dynamics of
these systems. Consequently, the concept
of higher-order differentiation and integration are widely  studied in diverse disciplines of
 science and engineering. For example, it is well understood that the first derivative
($\frac{d}{dt} f(t)$) and
second derivative ($\frac{d^{2}}{dt^{2}} f(t)$) of a function describe the rate of change and measure of concavity, respectively.
However, we rarely think what if the order ($n$) of higher-order derivative
 ($\frac{d^{n}}{dt^{n}}$) becomes a real, complex or an irrational number?
 One immediate question arises in our
 minds is the existence or possibility of such a concept in mathematics. Interestingly, this seemingly new concept
 dates back to 1695 when L'H\^{o}pital asked Leibniz regarding his notation $\frac{d^n y}{dx^n}$: ``{\em what if $n$ is $\frac{1}{2}$}''.
 In reply, Leibniz \cite{Le_1} prophesied in his letter,
 ``\ldots \textit{Thus it follows that ${d^{\frac{1}{2}} x}$ will be equal to $ x\sqrt{dx:x} $. This is an apparent paradox from which, one
 day, useful consequences can be drawn} \ldots''.
Leibniz's initial work
on the problem of defining the derivative of arbitrary order
gave birth to a new field of research in mathematics (called \textit{fractional calculus}) and
attracted the attention of many  physicists, engineers
and geometers. Some of the  great mathematicians
and physicists who touched the field of fractional calculus
are Riemann, Liouville, Laurent, Heaviside and  Riesz \cite{B2_93}.

 The concept of fractional calculus has great potential to change the way we model and analyze the systems.
 It provides good opportunity to scientists and engineers for revisiting the origins. We briefly outline some of the
 the main applications of fractional calculus in Table \ref{tab:applications}. The importance of fractional
calculus can be realized by the following  quote from Miller and Ross \cite{B2_93}. They stated:
 \begin{verse}

  \textit{``\ldots The  fractional  calculus  finds use  in many fields of science  and engineering,  including  fluid flow,  rheology,  diffusive  transport akin to diffusion,
electrical  networks,  electromagnetic  theory,  and probability\ldots. It  seems that  hardly a  field of science or
engineering has  remained  untouched by this topic \ldots" }
\end{verse}
\begin{table}
\begin{center} \label{tab:applications}
     \begin{tabular}{ | l | p{7cm} |}
     \hline
     \textbf{Field } & \textbf{Applications}  \\ \hline\hline
     Control Engineering   & - System identification \cite{sysid_99}\\
                           & - Biomimetic (bionics) control \cite{biom_04} \\
                           & - Trajectory control \cite{tr_2002}\\
                           & - Temperature control \cite{tmp_02}\\
                           & - Fractional $PI^{\alpha}$ controller \cite{pi_07}  \\ \hline
     Signal Processing  &  - Fractional order integrator   \cite{dspdiffint_08} \\
                        &  - Fractional order FIR differentiator \cite{fracFIR_01}\\
                        &  - IIR-type fractional order differentiator \cite{fracIIR_03} \\
                        &  - Modeling of speech signals \cite{speech_07}\\\hline
     Image Processing   & - Image restoration and edge detection \cite{image_03} \\
                        & - Satellite image classification \cite{sattliteimage_03} \\ \hline
     Electromagnetics   & - Fractional curl operators \cite{Engheta_98,Naqvi_04}\\
                        & - Fractional Rectangular waveguides \cite{faryad_07}\\ \hline
     Communication   & - Secure chaotic communication \cite{chaotic_09} \\
                     & - Informational network traffic modeling \cite{informational_01}\\ \hline
     Biology   & - Neuron modeling \cite{neuro_94}\\
               & - Biophysical processes \cite{biophy_09} \\
               & - Modeling of complex dynamics of tissues \cite{tissue_10} \\
               \hline\hline

 \end{tabular}
  \end{center}
  \caption{Applications of Fractional Calculus}
\end{table}

Nowadays engineering  systems exhibiting fractional order dynamics
are  increasingly used in some safety-critical applications such as
control systems, signal processing, electromagnetics and electrical networks (as listed in Table \ref{tab:applications}).
For example, fractional meta-materials based devices are used to
build sensitive military and defence equipments and electromagnetic stealth technology \cite{Matameterial}.
Considering these facts, it is quite  interesting and important to build a logical reasoning framework
which can be used to formally verify such sophisticated applications within the sound core
of a proof assistant. In fact, proof assistants have been  successfully
used to formalize and verify some  challenging and paradoxical mathematical results, e.g.,
the formal proofs of the Kepler Conjecture
(Flyspeck project) \cite{flyspeck}
and  the Odd Order Theorem \cite{ODD_ORDER}.

In this paper, we present details of an ongoing project\footnote{\url{http://save.seecs.nust.edu.pk/projects/fc.html}}  to develop a formal reasoning support for fractional calculus in higher-order-logic theorem prover. This project was originally started at the System Analysis and Verification
(SAVe) lab\footnote{\url{http://save.seecs.nust.edu.pk}} in 2010. Earlier formalization was done in the HOL4 theorem prover
with the main focus on  fractional operators for real-valued functions and the verification of
 fractional order electrical components. Later on, the scope of the project was expanded to
formalize fractional calculus involving complex-valued functions due to its various engineering applications (as listed in Table \ref{tab:applications}). Currently, we are using the
  HOL Light theorem prover due to the availability of rich multivariate analysis libraries
 including Harrison's recent formalization of complex-valued
 Gamma function\footnote{{\url{https://code.google.com/p/hol-light/source/browse/trunk/Multivariate/gamma.ml}}}
 as well as the interesting related projects like Flyspeck \cite{flyspeck} and  the formalization of  optics
 theories (i.e., ray, wave, electromagnetic and quantum) \cite{do-form-journal}.

The rest of the paper is organized as follows: In Section \ref{sec:fractional_calculus}, we briefly
review some commonly used notations and definitions of fractional order operators. We provide an outline of
the proposed formalization framework in Section \ref{sec:formal_analysis}. Consequently, the current status of the formalization
and future milestones are discussed in Section \ref{sec:formalization}. Finally, we conclude the paper in
Section \ref{sec:conclusion}.

\section{Mathematical Framework of  Fractional Calculus} \label{sec:fractional_calculus}
 There are different notations available for fractional derivatives and integrals.
 We use $J^{v}_{a}f(x)$ and $D^{v}f(x)$ for fractional integral and fractional derivative, respectively.
  In these notations, $v$ is the order of integration or differentiation and $a$ is the lower limit of integration.

 For every function ($f:\mathbb{C}\rightarrow\mathbb{C}$); and for every number $v \in \mathbb{R}$ or $\mathbb{C}$, $J^{v}_{a}$ and $D^{v}$ should be related to $f$ by the following criteria \cite{Das:2007:FFC:1564573}.

\begin{enumerate}
\item If $f(x)$ is an analytic function, then $J^{v}_{a}f(x)$ and $D^{v}f(x)$ must also be an analytic function of the variable $x$ and of the order $v$ of integration or differentiation.
\item The operations  $J^{v}_{a}f(x)$ and $D^{v}f(x)$ must produce the same result as ordinary integration/differentiation when $v$ is a positive integer.
\item The fractional operators must be linear.
\begin{equation}
 J^{v}_{a} [\alpha f(x) + \beta g(x)] = \alpha J^{v}_{a} f(x) + \beta J^{v}_{a} g(x)
\end{equation}
\begin{equation}
 D^{v} [\alpha f(x) + \beta g(x)] = \alpha D^{v} f(x) + \beta D^{v} g(x)
\end{equation}

	\item The operation of order zero must leave the function unchanged.
\begin{equation}
  J^{0}_{a} f = f \texttt{\ \ \ \ and \ \ \  }  D^{0} f = f
\end{equation}
 	\item The law of exponents must hold for integration and differentiation  of arbitrary order under sufficient conditions on function $f$.
 \begin{equation}
  J^{u}_{a}( J^{v}_{a}f)=  J^{u + v}_{a} f \texttt{\ \ and \ \  }  D^{u}(D^{v}f)=  D^{u + v} f
\end{equation}
   \end{enumerate}

\noindent Fractional integrals and fractional derivatives are also referred to as  Differintegrals \cite{B1_74} and  there are more than ten
well-known definitions for Differintegrals \cite{APP_10}. We describe here two of them,  which are  most widely used in analyzing real-world problems. These are the Riemann-Liouville and Gr\"{u}nwald-Letnikov definitions, which are also equivalent for a wide class of functions \cite{c_2}.

 \subsection*{\textbf{Riemann-Liouville (RL) Definition:}}

\begin{equation}\label{fractional_integration}
J^{v}_{a}f(x) = \frac{1}{\Gamma (v)}\int_{a}^{x} (x-t)^{v-1}f(t)dt
\end{equation}

\noindent where $J^{v}_{a}f(x)$ represents  fractional integration with order  $v$ and  lower integration limit $a$. The parameter $a = 0$ gives the Riemann definition and
 $a = -\infty$ gives the Liouville definition of fractional integration. Indeed Equation (\ref{fractional_integration})
 is the generalization  of Cauchy's repeated integration formula to non-integer $v$ \cite{BROSS_75}.
 Where  $\Gamma$ (.) in the above definition denotes the Gamma function which is defined using the well-known improper integral as follows:
\begin{equation}\label{gamma_improper}
\Gamma(z)= \int_{0}^{\infty} t^{z-1}e^{-t}dt
\end{equation}
\noindent for $Re(z) > 0$.

The fractional differentiation is given as follows:

  \begin{equation}\label{fractional_diff}
D^{v}f(x) = (\frac{d}{dx})^{m} J^{m-v}_{a}f(x)
     \end{equation}

\noindent where $m$ represents the ceiling of $v$, i.e., $\lceil v \rceil$.

 \subsection*{\textbf{Gr\"{u}nwald-Letnikov (GL) Definition:}}

\begin{equation}
_{c}D_{x}^{v}f(x)= \lim_{h\rightarrow0}h^{-v}\sum_{k=0}^{[\frac{x-c}{h}]}(-1)^{k}{v\choose k} f(x - kh)
\end{equation}

\noindent Gr\"{u}nwald-Letnikov definition  caters for both  fractional differentiation and integration, as positive values of $v$ give fractional differentiation and negative values of $v$ give fractional integration. Here, ${v\choose k}$ represents the binomial coefficients, which are described in terms of the Gamma function.


\section{Formal Analysis Framework} \label{sec:formal_analysis}

\begin{figure}[h]
  \centering
{\includegraphics[width=12.2cm]{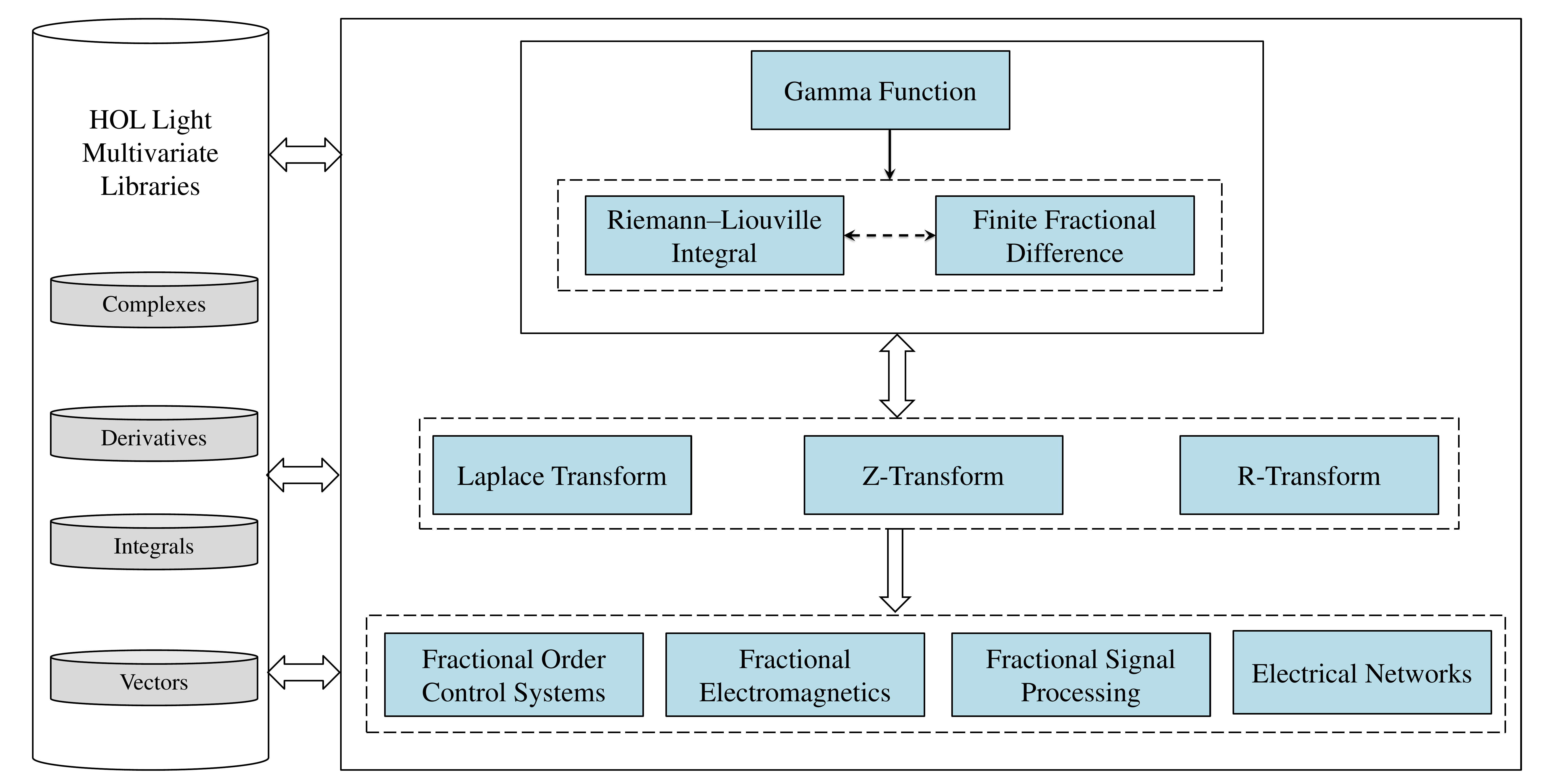}}
\caption{Formalization Framework for Fractional Calculus}
\label{fig:proposed} 
\end{figure}

The proposed framework, given in Figure \ref{fig:proposed}, outlines the main ideas and roadmap to
formalize the basic theory behind fractional calculus.
The whole framework can be decomposed mainly into three major parts which are the
formalization of the core definitions of fractional order operators, formalization of
supporting transformations (i.e., Laplace transform \cite{ogata_modern}, Z-transform \cite{DSP_OPENHEIUM}  and R-Transform \cite{R-transform}) and
engineering applications. The first part heavily relies upon the Gamma function
as mentioned in Section \ref{sec:fractional_calculus}. So the core step is to formalize the
Gamma function in higher-order logic (HOL) and verify its important properties. Consequently, any definition of
fractional order operators can be formalized in HOL. However, our focus is two main definitions, i.e.,
Riemann-Liouville (RL) and fractional difference which indeed represent continuous and discrete versions of
fractional order operators, respectively.  This step also involves the validation of all the properties mentioned
in Section \ref{sec:fractional_calculus}. This requires some important results of multivariate
calculus such as the notion of Lebesgue measurability and Fubini's theorem which provides the reasoning support for
iterated and double integrals. Interestingly, both of these requirements are available in the multivariate
analysis libraries of HOL Light. The second step is the formalization of important integral transforms which are necessary to
analytically solve linear fractional differential and difference equations. We mainly focus on three transforms, namely the Laplace
 transform, the Z-transform, and the recently introduced R-transform \cite{R_TRANSFORM}. All of these transformations
are used to transform complicated fractional differential (or difference) equations to algebraic equations which
are easier to manipulate and to deduce interesting properties.
Building upon these fundamentals, our ultimate goal is to formally verify a variety of
engineering systems including  control systems, signal processing, electromagnetics and
electrical networks. All formalization steps make use of different multivariate theories of
HOL Light, e.g., derivatives, integrals, complex vectors and measure spaces. Finally,
the developed libraries of this project will become part of the existing HOL Light libraries.

\section{Current Status and Future Milestones} \label{sec:formalization}
As mentioned earlier, the project initially considered only real-order fractional operators in the HOL4 theorem prover.
The main difficulty was the little  support to handle improper integrals which was required to formalize the
real-valued Gamma function. Therefore, we extended the  integration theory of HOL4 by formalizing a variant of
improper integrals using sequential limits. This was then used to formalize the
Gamma function and verify some of its
main properties, such as the pseudo-recurrence relation ($\Gamma(z+1)=z\Gamma(z)$), the functional equation ($\Gamma(1)=1$) and the factorial generalization ($\Gamma(k+1)=k!$) \cite{Umair_Gamma}.
We utilized these foundations to
formalize  Differintegrals, given in Equations (\ref{fractional_integration}) and (\ref{fractional_diff}),
which in turn can be used to represent the dynamics of fractional order systems in higher-order logic.
We also verified theorems corresponding to some commonly used properties of  Differintegrals namely
Identity and Linearity. Consequently, we conducted the formal analysis of a
 fractional order electrical component namely  resistoductor, a fractional integrator and a fractional differentiator circuit \cite{Umair_FMCAD_11}. Later on, the scope of the project was revised to include
complex-valued  functions and complex order fractional operators in HOL Light. The main requirement was to formalize the
complex-valued Gamma function, Laplace and Z-transforms.
However, the Gamma function\footnote{{\url{https://code.google.com/p/hol-light/source/browse/trunk/Multivariate/gamma.ml}}}
was formalized by Harrison in early 2014. In the meantime, we formalized the basic theories of the Laplace transform \cite {laplace_lpar}
and the Z-transform \cite{umair_itp_z}. Currently, we are working on three main topics which include: 1) formal proofs of the
uniqueness of Laplace and Z-transforms which are required to formally verify the inverses of these transforms; 2) vectorial Z-transform,
which extend the simple Z-transform over complex vectors; and 3) fractional difference equations, which are mainly based
on Gamma function, infinite summations and products over complex functions.
Finally, we outline the major tasks to  achieve the future milestones as follows:
\begin{itemize}
                        \item Formalization of R-Transform.
                        \item Formalization of Differintegrals for complex-valued functions. This is mainly the generalization of the formalization which was developed in HOL4.
                        \item Formalization of linear fractional differential and difference equations with support to analytical
                        solutions using the transform methods.
                      \end{itemize}

During the course of this project, two master and two PhD students have contributed to the formalization.
Interestingly, all of them are mainly electrical engineers without prior background of formal methods and
higher-order-logic theorem proving. Given the complexity and interdisciplinarity of this research project, it
is quite encouraging
to see people with an engineering (or physics) background
to use proof assistants as a complementary tool.
The formalization of the fractional calculus is quite challenging
 as it requires advanced mathematical concepts of vector
 integration and Lebesgue measurable functions, etc. So expertise in
 formal reasoning about these complex mathematical phenomena
is required for this formalization, which is quite
 unique compared to reasoning about software and digital hardware systems.
The learning curve of HOL Light varies from student to student.
 Generally, students start proving basic math equations after a
 couple of months and the pace of formalization increases over time.
 Learning HOL Light libraries is not difficult once the basic
 concepts have been grasped by the user.
 The formalization of the improper integrals, the Gamma function, the fractional
 calculus, the Z-transform and the Laplace transform is approximately 15,000 lines of HOL Light code.
 One of the major obstacles in the formalization was the identification of suitable
 mathematical definitions and models.
Sometimes, textbook proofs do not follow due to various reasons
 (corner cases, or the proof steps are too abstract, etc.) and they needed
 to be re-proved on paper with subtle details.
Consequently, we have to modify the definitions and thus change the proofs.
 But now the current formalization seems quite stable as most of the classical properties
 have been formally verified for our definitions. We believe that future
 developments can be built on the foundations that have been formalized
 as most of the work is for general systems. Finally, another important aspect of this project is
the potential to apply developed theories to various applications other than fractional calculus. For example, we
demonstrated the use of the Gamma function in probability theory \cite{Umair_Gamma},
the Z-transform in signal processing \cite{umair_itp_z}, and the Laplace transform in power electronics \cite{laplace_lpar}.

\section{Conclusion} \label{sec:conclusion}
In this paper, we mainly presented the motivation and ongoing activities of
our long term project about the formalization of fractional calculus in the
HOL Light theorem prover.
The main contribution of this project is a comprehensive framework of formal definitions
and theorems about fractional calculus which can be used
to verify modern control, signal processing and electromagnetic systems.
Some future directions and recommendations for HOL Light are
 the improvements in the  visualization of
 proofs, better automation and more accessible tutorials with examples
 from different engineering/physics topics.

\bibliographystyle{plain}
\bibliography{biblio}

\begin{thebibliography}{10}

\bibitem{chaotic_09}
N.~Pariz A.~Kiani-B, K.~Fallahi and H.~Leung.
\newblock A {C}haotic {S}ecure {C}ommunication {S}cheme {U}sing {F}ractional
  {C}haotic {S}ystems {B}ased on an {E}xtended {F}ractional {K}alman {F}ilter.
\newblock {\em Communications in Nonlinear Science and Numerical Simulation},
  14:863--879, 2009.

\bibitem{do-form-journal}
S.~K. Afshar, U.~Siddique, M.~Y. Mahmoud, V.~Aravantinos, O.~Seddiki, O.~Hasan,
  and S.~Tahar.
\newblock {Formal Analysis of Optical Systems}.
\newblock {\em Mathematics in Computer Science}, 8(1):39--70, 2014.

\bibitem{R-transform}
F.~M. Atici.
\newblock {A Transform Method in Discrete Fractional Calculus}.
\newblock {\em International Journal of Difference Equations}, 2(2):165--176,
  2007.

\bibitem{R_TRANSFORM}
F.~M. Atici and P.~W. Eloe.
\newblock {I}nitial {V}alue {P}roblems in {D}iscrete {F}ractional {C}alculus.
\newblock {\em Proceeding of the American Mathematical Society},
  137(3):981--989, 2009.

\bibitem{neuro_94}
T.~J. Auastasio.
\newblock {The Fractional-Order Dynamics of Brainstem Vestibulo-Oculomotor
  Neurons }.
\newblock {\em Biological Cybernetics}, 72(1):69--79, 1994.

\bibitem{sattliteimage_03}
E.~Cuestab C.~Quintanoa.
\newblock {Improving Satellite Image Classification by Using Fractional Type
  Convolution Filtering}.
\newblock {\em International Journal of Applied Earth Observation and
  Geoinformation}, 12(4):298--301, 2010.

\bibitem{biom_04}
Y.~Q. Chen, D.~Xue, and H.~Dou.
\newblock {Fractional Calculus and Biomimetic Control}.
\newblock In {\em Robotics and Biomimetics}, pages 901 --906. IEEE, 2004.

\bibitem{biophy_09}
M.~Mar\'{\i}n D.~M.~Dom\'{\i}nguez and M.~Camacho.
\newblock {Macrophage Ion Currents are Fit by a Fractional Model and Therefore
  are a Time Series with Memory }.
\newblock {\em European Biophysics Journal}, 38(4):457--464, 2009.

\bibitem{APP_10}
M.~Dalir and M.~Bashour.
\newblock Application of {F}ractional {C}alculus.
\newblock {\em Applications of Fractional Calculus in Physics}, 4(21):12, 2010.

\bibitem{Das:2007:FFC:1564573}
S.~Das.
\newblock {\em Functional {F}ractional {C}alculus for {S}ystem {I}dentification
  and {C}ontrols}.
\newblock Springer, 2007.

\bibitem{tr_2002}
Fernando B.~M. Duarte and Jos{\'e} Ant{\'o}nio~Tenreiro Machado.
\newblock {Pseudoinverse Trajectory Control of Redundant Manipulators: A
  Fractional Calculus Perspective}.
\newblock In {\em International Conference on Robotics and Automation}, pages
  2406--2411. IEEE, 2002.

\bibitem{diff_eq_book}
S.~Elaydi.
\newblock {\em {An Introduction to Difference Equations}}.
\newblock Springer, 2005.

\bibitem{Engheta_98}
N.~Engheta.
\newblock {Fractional Curl Operator in Electromagnetics}.
\newblock {\em Microwave Optics Technology Letters}, 17(2):86--91, 1998.

\bibitem{faryad_07}
M.~Faryad and Q.~A. Naqvi.
\newblock Fractional {R}ectangular {W}aveguide.
\newblock {\em Progress In Electromagnetics Research, PIER}, 75:383--396, 2007.

\bibitem{ODD_ORDER}
G.~Gonthier, A.~Asperti, J.~Avigad, Y.~Bertot, C.~Cohen, F.~Garillot,
  S.~Le~Roux, A.~Mahboubi, R.~OConnor, S.~Ould~Biha, I.~Pasca, L.~Rideau,
  A.~Solovyev, E.~Tassi, and L.~Théry.
\newblock {A Machine-Checked Proof of the Odd Order Theorem}.
\newblock In {\em Interactive Theorem Proving}, volume 7998 of {\em LNCS},
  pages 163--179. Springer, 2013.

\bibitem{flyspeck}
T.~C. Hales.
\newblock {Introduction to the Flyspeck Project}.
\newblock In {\em Mathematics, Algorithms, Proofs}, volume 05021 of {\em
  Dagstuhl Seminar Proceedings}, pages 1--11, 2005.

\bibitem{sysid_99}
T.~T. Hartley and C.~F. Lorenzo.
\newblock Fractional {S}ystem {I}dentification: {A}n {A}pproach {U}sing
  {C}ontinuous {O}rder {D}istributions.
\newblock Technical report, National Aeronautics and Space Administration,
  Glenn Research Cente NASA TM, 1999.

\bibitem{speech_07}
W.M.~Ahmad K.~Assaleh.
\newblock Modeling of {S}peech {S}ignals {U}sing {F}ractional {C}alculus.
\newblock In {\em International Symposium on Signal Processing and its
  Applications}, pages 1--4. IEEE, 2007.

\bibitem{dspdiffint_08}
B.~T. Krishna and K.~V. V.~S. Reddy.
\newblock Design of {D}igital {D}ifferentiators and {I}ntegrators of {O}rder
  $\frac{1}{2}$.
\newblock {\em World Journal of Modelling and Simulation}, 4:182--187, 2008.

\bibitem{Le_1}
G.~W. Leibnitz.
\newblock Leibnitzen's {M}athematische {S}chriften.
\newblock {\em SIGDA News Letter}, 2:301--302, 1962.

\bibitem{Matameterial}
K.~A. Lurie.
\newblock {\em {An Introduction to the Mathematical Theory of Dynamic
  Materials}}.
\newblock Springer, 2007.

\bibitem{tissue_10}
Richard~L. Magin.
\newblock Fractional {C}alculus {M}odels of {C}omplex {Dy}namics in
  {B}iological {T}issues.
\newblock {\em Computers and Mathematics with Applications}, 59:1586--1593,
  2010.

\bibitem{pi_07}
G.~Maione and P.~Lino.
\newblock New {T}uning {R}ules for {F}ractional $pi^{\alpha}$ {C}ontrollers.
\newblock {\em Nonlinear Dynamics}, 49(1-2):pp 251--257, 2007.

\bibitem{B2_93}
K{.}~S{.} Miller and B{.} Ross.
\newblock {\em An Introduction to Fractional Calculus and Fractional
  Differential Equations}.
\newblock John Willey, 1993.

\bibitem{Naqvi_04}
Q.~A. Naqvi and M.~Abbas.
\newblock {Complex and Higher Order Fractional Curl Operator in
  Electromagnetics }.
\newblock {\em Optics Communications}, 241:349--355, 2004.

\bibitem{ogata_modern}
K.~Ogata.
\newblock {\em Modern Control Engineering}.
\newblock Prentice Hall, 2010.

\bibitem{B1_74}
K{.}~B{.} Oldham and J{.} Spanier.
\newblock {\em The {F}ractional {C}alculus}.
\newblock New York, Academic Press, 1974.

\bibitem{DSP_OPENHEIUM}
A.~V. Oppenheim, R.~W. Schafer, and J.~R. Buck.
\newblock {\em {Discrete-Time Signal Processing}}.
\newblock Prentice Hall, 1999.

\bibitem{image_03}
B.~Mathieu{,} P. Melchior{,}~A. Oustaloup and Ch. Ceyral.
\newblock Fractional {D}ifferentiation for {E}dge {D}etection.
\newblock {\em Signal Processing}, 83:2421--2432, 2003.

\bibitem{tmp_02}
I.~Petr\'{a}s and B.M. Vinagre.
\newblock {Practical Application of Digital Fractional-Order Controller to
  Temperature Control}.
\newblock {\em Acta Montanistica Slovaca}, 7(2):131--137, 2002.

\bibitem{c_2}
I.~Podlubny.
\newblock {F}ractional {D}ifferential {E}quations, {A}cademic {P}ress.
\newblock 1999.

\bibitem{BROSS_75}
B.~Ross.
\newblock A {B}rief {H}istory {A}nd {E}xposition of {T}he {F}undamental
  {T}heory of {F}ractional {C}alculus.
\newblock In {\em Fractional Calculus and Its Applications}, volume 457 of {\em
  Lecture Notes in Mathematics}, pages 1--36. Springer, 1975.

\bibitem{Umair_FMCAD_11}
U.~Siddique and O.~Hasan.
\newblock Formal {A}nalysis of {F}ractional {O}rder {S}ystems in {HOL}.
\newblock In {\em Formal Methods in Computer Aided Design}, pages 163--170.
  IEEE, 2011.

\bibitem{Umair_Gamma}
U.~Siddique and O.~Hasan.
\newblock {On the Formalization of Gamma Function in HOL}.
\newblock {\em Journal of Automated Reasoning}, 53(4):407--429, 2014.

\bibitem{umair_itp_z}
U.~Siddique, M.~Y. Mahmoud, and S.~Tahar.
\newblock {On the Formalization of Z-Transform in HOL}.
\newblock In {\em Interactive Theorem Proving}, volume 8558 of {\em LNCS},
  pages 483--498. Springer, 2014.

\bibitem{laplace_lpar}
S.~H. Taqdees and O.~Hasan.
\newblock {Formalization of Laplace Transform Using the Multivariable Calculus
  Theory of HOL-Light}.
\newblock In {\em Logic for Programming, Artificial Intelligence, and
  Reasoning}, volume 8312 of {\em LNCS}, pages 744--758. 2013.

\bibitem{fracFIR_01}
C.~C. Tseng.
\newblock {Design of Fractional Order Digital FIR Differentiators}.
\newblock {\em IEEE Signal processing Letters}, 8(3):77--79, 2001.

\bibitem{fracIIR_03}
B.~M.~Vinagreb Y.~Q.~Chena.
\newblock {Fractional Differentiation for Edge Detection}.
\newblock {\em Signal Processing}, 83:2359--2365, 2003.

\bibitem{math_modeling}
X.~S. Yang.
\newblock {\em {Mathematical Modeling with Multidisciplinary Applications}}.
\newblock John Wiley, 2013.

\bibitem{informational_01}
V.~Zaborovsky and R.~Meylanov.
\newblock {Informational Network Traffic Model Based on Fractional Calculus }.
\newblock In {\em Proceedings of the International Conference Info-tech and
  Info-net}, pages 58--63. IEEE, 2001.

\end{thebibliography}

\end{document}